# Firm's information environment and stock liquidity: Evidence from Tunisian context

Nadia LOUKIL[1] and Ouidad YOUSFI[2]

**Abstract**

This paper analyzes the relationship between public disclosure, private information and stock liquidity in Tunisian context using a sample of 41 listed firms in the Tunis Stock Exchange in 2007. First, we find no evidence that there is a relation between public and private information. Second, Tunisian investors do not trust the information disclosed in both annual reports and web sites, consequently it has no effects on stock liquidity, in contrast with private information.

---

[1] Nadia LOUKIL is associated at Fiesta in the ISG of Tunis. Corresponding Email: nadialoukil@gmail.com. Address : 293 Cité El Hana Bizerte 7000, Tunisia
[2] Ouidad YOUSFI is an assistant Professor at the Department of Accounting and Financial Methods in the ESSEC de Tunis and is associated at Fiesta in the ISG de Tunis (Tunisia) and EconomiX in the Université de Paris West Nanterre la Défense (France).

# 1. Introduction

It is commonly known that asymmetric information problems lead to agency conflicts between managers and outside investors, which decrease both the volume and number of transactions in capital market (Akerlof, 1970). Voluntary disclosure of information has double role. First, it is a way to mitigate information asymmetry and consequently agency conflicts; investors use disclosed information to choose profitable projects. Second, even when the project is carried out, information disclosure deters also managers from opportunistic behavior. For example, they cannot take excessive risky decisions to expropriate investors' wealth (Bushman and Smith, 2003). Moreover, Myers and Majluf (1984) argue that disclosure reduces agency costs and even the cost of outside financing (the pecking order theory). Since, "good" managers will be encouraged to disclose more voluntary information, which is, in turn, considered as a good signal of the quality of corporate governance (Chen et *al*., 2007).

In the last years, corporate information disclosure has received considerable attention: many empirical studies argue that voluntarily disclosure in developed markets which are strongly regulated reduces capital cost[3] and improves stock liquidity[4]. However, in emerging markets, empirical studies are divided and their results are not conclusive: Hassan et al. (2009), Wang et al. (2008) and Chen et *al*. (2009) show that there is no significant effect of information disclosure neither on firm's value nor on financing cost (debt and equity). Gana and Chemli (2008) find that stock liquidity decreases with the level of information disclosure. However, Mattoussi et al. (2004) and Haddad et *al.* (2009) find a positive relationship between stock liquidity and disclosure level. Notice that the main source of information in these studies is the information publicly and voluntarily disclosed in annual reports. However, the firm's information environment consists of public information (disclosed in annual reports, web sites and conference calls) and private information disseminated through informed trading. Financial literature on information disclosure argues that public and private information are substitutes (Verrecchia, 1982 and Diamond, 1985) or complement (Kim and Verrecchia, 1991 and McNichols and Trueman, 1994).

In this paper, we raise the question: *what is the impact of the informational environment on stock liquidity in the Tunisian market?*

---

[3] See among others, Botosan and Plumlee (2002) and Hail (2002).

[4] See among others Welker (1995), Heflin et *al.*( 2005 and 2007) Brown and Hillegeist (2007) and Chen et *al.* (2007).

To answer this question, we examine how Tunisian investors make investment decisions and if there is a relationship between public and private information. In contrast with Mattousi et al. (2004) and Gana and Chemli (2008) who were interested only in public information disclosed in annual reports, we consider both public information (in annual reports and web sites) and private information (order flows).

In Tunisia, the law No 94-117 has fixed mandatory information, the conditions and the timing of disclosure. In addition, the Financial Market Council (*Conseil des Marchés Financiers*, CMF) set some rules about the kind of information that should be disclosed in annual reports. In 2008, CMF reformed presented a detailed reference model of annual reports. However, it does not supervise the content of annual reports and does not impose punishment if firms did not publish mandatory information. Consequently, Tunisian firms are not constrained to disclose more information. Fitch Ratings (2009) argues that Tunisian market is poorly regulated and has transparency problems. During the period 2006-2009, Tunisian market was assigned a disclosure index equal to 0 by the Doing Business reports[5].

The current study has three main results. First, we find no relationship between public and private information. This finding implies that public information does not reduce adverse selection problems as signaling theory predicts. Second, in contrast with prior Tunisian studies (Mattoussi and al., 2004; Gana and Chemli, 2008), we find that voluntary information disclosure in annual reports and on websites has no significant effect on stock liquidity. Our results show that information disclosure does not solve information asymmetry and that Tunisian investors do not rely on it to make their investment decisions. Third, it seems that Tunisian investors are overconfident: they rely only on their private information even when there is an arrival of new flow of public information. They did not update their beliefs and traded aggressively (Daniel et al., 1998) which may decrease stock liquidity.

The remainder of this paper is organized as follows. Section 2 reviews the literature and provides hypothesis. The sample and the methodology are presented in Section 3. Results are discussed in Section 4. In Section 5, we test the robustness of our results. We conclude in Section 6.

---

[5] The extent of disclosure index is yearly index provided by Doing Business. It varies between 0 and 10, with higher values reflecting greater disclosure. This index includes 5 components: 1)What corporate body provides legally sufficient approval for the transaction; 2) Immediate disclosure to the public and/or shareholders;3) Disclosures in published periodic filings; 4) Disclosures by controlling shareholder to board of directors; 5) Requirement that an external body review the transaction before it takes place.

## 2. Survey of the literature

Adverse selection models (see among other Bagehot, 1971; Kyle 1985 and Glosten and Milgrom, 1985) are based on the assumption that market makers establish a large spread to minimize potential losses due to informed trading and simultaneously to maximize potential gains due to uninformed trading. Hence, adverse selection risk induces a high cost of transaction. Even without market makers, as in order-driven market, it is shown that adverse selection problems have effects on trading process and stock liquidity. Handa and Schwartz (1996) show that liquidity suppliers, who can be considered as market makers, prefer doing few orders to compensate the losses of informed trading with the gains of uninformed trading. Later, Handa et al. (2003) find that high spread is explained by adverse selection problems. In order to diminish information asymmetry, traders prefer public information rather than private information: collecting private information is too costly. This is supposed to improve the market conditions: homogenize investors' opinions and reduce speculative positions (Verrecchia, 1982; Diamond, 1985). More information disclosure is supposed to reduce asymmetric information which reduces transactions costs and improves stock liquidity (Amihud and Mendelson, 1986, and Diamond and Verrecchia, 1991). Furthermore, Brown and Hillegeist (2007) point out that information disclosure improves the firm's image in its market. In contrast with traders, investors rely on private information particularly when they are expecting future public disclosures in pre-announcement period. So, public disclosures may make asymmetric information problems more severe (Kim and Verrecchia 1991; McNichols and Trueman, 1994).

Notice that most of the previous studies were conducted in developed markets, particularly in the American market. They show that improving disclosure leads to a decrease of the spread (Welker, 1995; Healy et al. 1999; Heflin et al. 2005; Chen et al. 2007) and consequently the compensation of market makers. In such conditions, they noticed that many market makers leave their job because of the low quoted depth (Heflin et al., 2005). These results are also consistent with those of Brown and Hillegeist (2007), who find that the disclosure's quality is negatively related to the level of information asymmetry. In other stock markets, Madrid stock exchange for example, Espinosa et *al.* (2008) join previous papers and highlight the positive relation between stock liquidity and disclosure level. In contrast, in Jordanian stock market, Haddad et al. (2009) find a negative relation between spread and disclosure level, but lower than in developed countries. Mattoussi et *al.* (2004) test the relationship between disclosure level and stock liquidity using a Tunisian data in 2001. Their results show that high level of disclosure diminishes quoted spread and increases quoted depth. Later, Gana and Chemli (2008) study the impact of disclosure level on spread using a sample of listed firms in the period 2001-2004. In contrast with Mattousi et *al.*, they show a positive and significant effect of voluntary disclosure on spread. One explanation is that Tunisian investors have no confidence on the disclosed information in annual reports and do not use it to make their decisions.

Thus, private information can increase adverse selection problems and lead consequently to a decrease of stock liquidity. In contrast, public/voluntary information disclosure may solve adverse selection problems and discourage collecting private information and improves stock liquidity. Based on these findings, we state two hypotheses:

*H1: Under asymmetric information, private information has negative effect on stock liquidity.*

*H2: The more public information is disclosed, the more stock liquidity will increase.*

## 3. Data and methodology
### 3.1. Data

First, we consider common stocks of firms listed in the Tunisian Stock Exchange in 2007. The initial sample contains 50 firms in both financial and non-financial sectors. Second, we eliminate stocks recently introduced in 2007 and non common ordinary stocks. Third, we exclude firms missing annual reports. Hence, 41 firms remain in our sample. Annual reports of these firms were provided by the CMF and brokerage firms. In order to examine disclosure on websites, data were gathered from firms' websites. For other variables, the data about daily trading (like for example price, trading volume, best ask and best bid) are provided by the Tunisian Stock Exchange.

**Table 1. Sample composition**

| Initial sample | |
|---|---|
| Shares listed in 2007 | 50 |
| New listed shares | 2 |
| Non common ordinary shares | 2 |
|  | 46 |
| **Shares by industry** | |
| Banks | 10 |
| Other financial firms | 11 |
| Services | 7 |
| Manufacturing firms | 18 |
|  | |
| **Annual reports not available** | 5 |
| **Final sample** | 41 |

### 3.2. Liquidity measure

In contrast with prior empirical studies using one-dimensional liquidity measures such as spread or depth (Mattoussi et al. 2004; Gana and Chemli, 2008), we choose a multidimensional measure: the turnover-adjusted number of non trading days. According to Liu (2006), this measure captures three dimensions of liquidity: potential delay for executing an order, the cost and the quantity of transaction. The Liu's measure is the standardized turnover-adjusted number of zero daily trading volumes which is supposed to be more appropriate to assess liquidity risk than average spread and illiquidity ratio of Amihud (2002) when the sample includes shares with high trading activity and shares with low trading activity. Indeed, the average spread and Amihud's ratio cannot be calculated in non trading days, while Liu'measure includes the effect of non trading on liquidity risk.

$$LIUM = \left[ NoZV + \frac{1/TURN}{Deflator} \right] \times \frac{252}{NoTD}$$

where NoZR is the number of zero-volume trading days and NoTD is the total number of trading days in the market over the year. Because this number can vary from one year to another, the factor $\frac{252}{NoTD}$ is used to standardize it to 252 days (average number of trading days in one year) to make this measure comparable over time; Deflator[6] is chosen arbitrary for all stocks, such that:

$$0 < \frac{1/TURN}{Deflator} < 1$$

### 3.3. Voluntarily information disclosure level

In order to measure the level of information disclosure, previous studies advance that annual reports are more "informative" than short-term reports and other sources of information (Lang and Lundholm, 1993; Botosan 1997). However, these studies were conducted in developed economies and consequently cannot be automatically generalized to emerging economies. Moreover, in addition to annual reports, we consider another source of information: firms' websites. To our knowledge there is no empirical study analyzing the link between stock liquidity and public information disclosed in websites in emerging and developed markets. For this reason, we construct two indexes to measure public information disclosed in annual reports and in web sites.

In prior studies[7] on voluntary disclosure, two categories of disclosure index were used. First, the indexes constructed by specialized agencies, like indexes of Corporate Information Committee of the Financial Analysts Federation (FAF), the Association of Investment Management and Research Corporation Information Committee (AIMR) and Standard & Poor's (S&P). They contain all the information provided by firms: annual, half-yearly, quarterly and other written information and information about investors' relations. However, these agencies are dealing only with large firms. Second, the indexes constructed by studies such as indexes of Botosan (1997) Eng and Mak (2003) and Wang et al. (2008) to measure the level of disclosure only in annual reports. These indexes depend significantly on subjective criteria, for example the author's approach (Marston and Shrives, 1991). Because of the absence of such agencies in Tunisia, we considered extensions of these indexes more appropriate to measure information disclosure in annual reports and in websites.

*Disclosure level in annual reports (BOTS)*

The first index we consider is an extension of Botosan index (Botosan, 1997) and measures the volume of public information in annual reports. It captures five types of information: (1) background information like for example management's objectives, business strategy and principal products; (2) historical summaries of annual financial results; (3) non-financial information such as market share and average compensation per employee; (4) forecasted information such as forecast of profits and cash-flows, and (5) management discussion and analysis about yearly changes that are not contained in financial statements. This index, initially constructed for non financial American firms, was adapted by Mattoussi et al. (2004) in financial sector. In this paper, we readjusted this index according to the CMF regulation.

---
[6] We use a deflator of 3500 000 in constructing LIUM
[7] See among others Chen et al. (2007), Botosan (1997) and Eng and Mak (2003)

To construct this index, first, we fix preliminary information items which may be disclosed in a voluntarily way. Botosan (1997) choose items according to the recommendations of the Jenkins report (AICPA, 1995). The Botosan' scoring procedure consists on assigning one point for qualitative information and one additional point for quantitative information. The score is the total points given to the firm divided by the highest score. Many researchers construct their own indexes based upon this index for different institutional setting, such as: Gul and Leung (2004) for Hong Kong listed firms; Patelli and Prencipe (2007) for Italians firms; Alsaeed (2006) for Saudi firms, Mattoussi et al. (2004) and Gana and Chemli (2008) for Tunisian firms.

Second, we adapt these items to Tunisian firms. We compare preliminary items with required elements according to the CMF. We take the example of information about historical financial results: Tunisians firm must provide some statistics about performance variations during the past 5 years. Thus, information disclosed for a period longer than 5 years is called voluntary information. However, indexes used in previous studies set the threshold of 2 years (Mattoussi et al., 2004; Gana and Chemli, 2008).

Third, we adjust our index to the practices in Tunisian firms. We apply items list to annual reports to exclude irrelevant items, such as: 1) not disclosed by any firm and 2) disclosed by all firms. Our final index includes 36 items which contain general information (12 items); summary of historical financial results (3 items), non-financial information (5 items), forecasting information (7 items) and analysis and discussion of the management (9 items).

The final step is to test the reliability of this constructed index. For this we use Cronbach's coefficient alpha that is commonly used to assess the internal consistency. Cronbach's coefficient indicates that the disclosure index shows satisfying internal consistency (Cronbach Alpha = 0,650).

*Disclosure level in firm' websites (SWEB)*

We construct another index to evaluate the level of information contained in firms' websites. Disclosure level (the volume and the quality of information) in websites is difficult to assess since it cannot be measured directly (Cooke and Wallace, 1989). Using Internet to disclose information has become a common practice in many companies. It is a way among others to reduce disclosure's costs (Healy and Palepu, 2001). It provides valuable information to investors who would like to invest in the firm. The corporate governance's principles of the OECD (2004) and the guide of good governance practices in Tunisian companies edited in 2008 encourage the use of Internet. We scrutinize websites to identify the main information, which is supposed to be helpful in making decision process. We retain six kinds of information: financial information not included in annual reports, availability of downloadable annual reports, availability of downloadable annual reports of previous years, access to press releases; access to press articles such as interviews with CEO (some press articles are downloadable) and availability of corporate governance data. This helps us to assign an index to each website.

**Table 2. Frequency of items identified in Tunisian firms' site web**

|  | Frequency | Percentage |
|---|---|---|
| **Existence of a website** | 31 | 67% |
| **Financial information** | 9 | 19% |
| **Availability of annual report** | 10 | 22% |
| **Availability of annual reports of previous years** | 5 | 11% |
| **Access to press releases** | 6 | 13% |
| **Access press articles** | 2 | 43% |
| **Availability of governance data** | 6 | 13% |

Only 67% of our firms have web sites, such as Amen bank, ASSAD, BIAT and TUNISAIR. We observe that websites contain annual report of the current year and some additional financial information and press articles. We state that information addressed to investors and the firms' shareholders is fewer than information addressed to customers and suppliers. Scoring procedure is to assign one point for each available item. The web index is the sum of points divided by the highest index (7 points). The test of reliability reveal a good internal consistency (Cronbach Alpha = 0,738).

### 3.4. Private information production

We use the percentage of informed trading as a measure of private information production, the average absolute value of imbalance order (**AIMO**). Easley et *al.* (1996) argue that uninformed investors submit buying and selling orders with equal probabilities. However, informed ones submit more purchase orders if they receive positive information signal and more sales orders if they receive negative information signal. Therefore, the difference between the two kinds of orders measures the information asymmetry. Hmaied et *al.* (2006) find that Tunisian investors have different behavior. In fact, in contrast with sellers who use only public information, buyers use private information to decide. Thus, they conclude that buyers submit more orders than sellers. According to Aktas et *al.* (2007), the probability of informed trading **AIMO** can be measured by:

$$AIMO = |(QB - QS)/(QB + QS)|$$

where QB and QS represent respectively orders quantity of ask and bid.

### 3.5. Control variables

In the current study, we retain the following control variables: volatility (**VLAT**), firm size (**SIZE**) and the book to market (**BTMK**). To measure stock return volatility, we use the standard deviation of daily returns. It captures total risk of stocks. Most of the studies[8] find that stocks with high volatility are riskier and consequently less liquid. In contrast, Kyle (1985) and Admati and Pfleiderer (1988) advance that volatility is positively associated with stock liquidity. Indeed, informed traders attempt to hide their trading among transactions of liquidity traders' transactions, which induces high volatility and high liquidity. Hence, an increase of volatility increases the liquidity in the market.

Merton (1987), Stoll and Whaley (1990) and Amihud and Mendelson (1986) find that the firm size, defined by logarithm of the firm's capitalization, increases with stock market liquidity. According to Fama and French (1993), book to market ratio captures the firm's risk, and investors ask for high premium to compensate them for the risk of holding their stocks. This is why, we include book to market ratio in order to control the effect of firm risk on stock liquidity.

---
[8] See among others, Stoll (1978), Amihud and Mendelson (1980) and Ho and Stoll (1981)

**Table 3. Definition of variables**

| *Variables* | *Abbreviations* | *Indicators* | *Expected signs* |
|---|---|---|---|
| Liquidity | **LIUM** | The standardized turnover-adjusted number of zero daily trading volumes | |
| Characteristics of information environment | **BOTS** | Botosan index modified | (+) |
| | **SWEB** | Siteweb index | (+) |
| | **AIMO** | The average absolute value of imbalance order | (-) |
| Control variables | **SIZE** | Market value of equity | (+) |
| | **BTMK** | A ratio of the book value of a assets to its market value | (-) |
| | **VLAT** | Standard deviation of the daily stock returns | (-) |

## 4. Empirical findings and discussions

### 4.1. Descriptive statistics

The table 4 presents descriptive statistics. The Panel A shows that potential delay in executing an order is on average 56 days. There is a big deviation for this variable (67 days), which implies that our sample includes high and low frequently traded stocks. In addition, we notice that the parameter of the firm's size is on average 129 millions (Ms) of TND[9], which varies between 6 Ms TND and 784 Ms TND. One explanation of the high dispersion of the firm's size is that our sample contains 10 of the largest firms[10] (66% of market share).

**Table 4. Descriptive statistics**

| A. descriptive statistics of liquidity and other stock characteristics | | | | |
|---|---|---|---|---|
| | **LIUM** | **BTMK** | **SIZE (MD)** | **VLAT** |
| **N** | 46 | 46 | 46 | 46 |
| **Mean** | 56.791 | 0.800 | 129 | 0.032 |
| **Median** | 26.313 | 0.815 | 52 | 0.015 |
| **Std. Deviation** | 68.987 | 0.292 | 174 | 0.046 |
| **Skewness** | 1.477 | 0.720 | 2.323 | 2.741 |
| **Kurtosis** | 1.266 | 2.680 | 5.761 | 7.445 |
| **Minimum** | 0.001 | 0.256 | 6 | 0.005 |
| **Maximum** | 243.067 | 1.800 | 784 | 0.223 |

---

[9] 1TND≈0,69665 USD
[10] These firms are SFBT, Tunisair, BT, BIAT, BH, UBCI, ATB, STB, Attijari Bank and Amen Bank

> **Legend**:
> BTMK=book-to-market ratio; SIZE= market capitalisation; VLAT=standard deviation daily returns; LIUM= standardised turnover-adjusted number of zero daily trading volume;

| B. Descriptive statistics of information environment | | | | | | | | |
|---|---|---|---|---|---|---|---|---|
| | **IGEN** | **HIST** | **INFI** | **PREV** | **GEST** | **BOTS** | **SWEB** | **AIMO** |
| **N** | 41 | 41 | 41 | 41 | 41 | 41 | 46 | 46 |
| **Mean** | 0.089 | 0.004 | 0.038 | 0.026 | 0.079 | 0.222 | 0.211 | 46.342 |
| **Median** | 0.080 | 0.000 | 0.043 | 0.022 | 0.087 | 0.217 | 0.143 | 40.073 |
| **Standard deviation** | 0.027 | 0.011 | 0.023 | 0.029 | 0.032 | 0.058 | 0.225 | 18.562 |
| **Skewness** | 1.486 | 2.951 | -0.310 | 0.880 | -0.361 | 0.391 | 1.150 | 14.48 |
| **Kurtosis** | 2.119 | 8.052 | -0.532 | -0.089 | -0.510 | 0.303 | 0.684 | 15.07 |
| **Minimum** | 0.057 | 0.000 | 0.000 | 0.000 | 0.000 | 0.103 | 0.000 | 25.158 |
| **Maximum** | 0.174 | 0.043 | 0.087 | 0.109 | 0.130 | 0.348 | 0.857 | 99.998 |

**Legend**
IGEN= General information; HIST= summary historical financial results; INFI= non financial information; PREV=forecasting information; MANG= analysis and discussion of the management; BOTS= disclosure index of annual reports; SWEB =website disclosure index; AIMO= absolute value of imbalance order.

The average stock return volatility is high (3%) and it varies between 22,3% and 0,5%. The ratio of book to market is on average 80%, which implies that stocks are overvalued. Statistics indicates that the average percentage of informed trading, which captures private information, is high (46%) and may reach a maximum level of 99%. The average disclosure index of annual reports is 22,2%, and varies between 10,3% and 34,8%. Its average deviation is 5,8%. This means that the level of information disclosure does not vary significantly among Tunisian firms. They prefer disclosing more information about management objectives, business strategy and the change of management activity, and some financial and forecasting information. The disclosure index on web sites is on average 21,1% and its deviation is 22,5%, which suggests that the content of websites varies significantly from one firm to another: (1) 15 firms have no sites, (2) 14 firms use websites as customer interface and do not disclose any information to investors[11], and (3) 17 firms disclose useful information for investors.

### 4.2. Correlation analysis

Table 5 reports that liquidity does not depend on voluntary disclosure in annual reports and in websites. In contrast, the percentage of informed trading is positively correlated with the timing of executing an order. This implies that stock liquidity is decreasing with private information. The information environment proxies (disclosure index on annual reports, disclosure index on websites and imbalance order) are not correlated between them, which is not consistent with the assumption that private and public information are not related.

---

[11] Tunisair, SIAME, Elmazraa and UBCI.

**Table 5. Spearman's correlation between informational environment of firms and control variables**

|      | BOTS  | SWEB    | AIMO   | LIUM     | BTMK   | VLAT   | SIZE  |
|------|-------|---------|--------|----------|--------|--------|-------|
| **BOTS** | 1,000 |         |        |          |        |        |       |
| **SWEB** | 0.071 | 1,000   |        |          |        |        |       |
| **AIMO** | 0.029 | -0.022  | 1,000  |          |        |        |       |
| **LIUM** | -0.077| -0.226  | 0.829**| 1,000    |        |        |       |
| **BTMK** | -0.127| 0.163   | -0.013 | -0.056   | 1,000  |        |       |
| **VLAT** | -0.079| 0.043   | -0.201 | -0.162   | 0.013  | 1,000  |       |
| **SIZE** | 0.071 | 0.421** | -0.264 | -0.406** | -0.271 | -0.176 | 1,000 |
| **Legend:** |||||||| 
| BTMK=book-to-market ratio; SIZE= market capitalization; VLAT=standard deviation daily returns; LIUM= standardized turnover-adjusted number of zero daily trading volume; BOTS= disclosure index on annual reports; SWEB= disclosure index on website; AIMO=absolute value of imbalance order.*, **: statistically significant for the threshold values of 5% and 1% respectively. |||||||

In fact, the policy of Tunisian firms in terms of corporate information disclosure does not add valuable and reliable information to investors; consequently, they do not rely on such information to make decisions. One explanation is that usually Tunisian investors did not rely on disclosed information: they prefer traditional ways to collect the information they need. Indeed, Dellagi et al. (2001) advance that Tunisians invest based on information provided by friends and relatives. Some of them are suspicious and do not trust these reports. We notice a positive correlation between web sites disclosure and the firm's size. This result shows that only the largest firms, particularly banks, disclose information through their websites.

### 4.3. Regression analysis

Hereafter, we test the following model to study the relation between stock liquidity and the variables presented above:

$$LIUM_i = \delta_0 + \delta_1 AIMO_i + \delta_2 BOTS_i + \delta_3 SWEB_i + \delta_4 VLAT_i + \delta_5 BTMK_i + \delta_6 SIZE_i + \varepsilon_i$$

**Table 6. Relationship between firms' information environment and stock liquidity**

|          | LIUM      |
|----------|-----------|
| **AIMO** | 3.296     |
|          | (6.73)**  |
| **BOTS** | -89.922   |
|          | (0.98)    |
| **SWEB** | -39.697   |
|          | (1.92)    |
| **VLAT** | 1.296     |
|          | (0.24)    |
| **SIZE** | -2.430    |
|          | (0.55)    |
| **BTMK** | 28.667    |
|          | (1.39)    |
| **Constant** | -40.623 |
|          | (0.44)    |
| **Observations** | 41 |
| **Adjusted R2** | 0.79 |
| **Legend** | |
| BOTS= disclosure index on annual reports; SWEB= disclosure index on website; AIMO=absolute value of imbalance order; BTMK=book-to-market ratio; SIZE= market capitalization; VLAT=standard deviation daily returns. | |
| *, **: statistically significant for the threshold values of 5%, and 1% respectively. | |

The model's estimation leads the validation of H1 and the rejection of H2. Indeed, we report a positive effect of percentage of informed trading on Liu measure. Then we conclude that private information increases adverse selection risk which reduces stock liquidity. In addition, we show that the level of information disclosure, measured by Botosan index has no effect on liquidity. This result is consistent with Hassan et al. (2009), who find no effect of voluntary information disclosure on Egyptian firms' value.

However, our study provides different results from Gana and Chemli (2008) and Mattoussi et *al.* (2004). For instance, Mattoussi et al. study demonstrates that information voluntarily disclosed in annual reports reduces information asymmetry and improves stock liquidity, in contrast with Gana and Chemli find the opposite effect. One explanation of their different findings is the use of different research method. They examine different periods using dynamic and static approaches. In contrast with indexes they used, we constructed a new index more appropriate to the Tunisian framework. Indeed, we have excluded from the Botosan index used by Mattoussi et al. (2004) and Gana and Chemli (2008) some items that are considered as mandatory items according CMF regulation. Hence, the significant effect of disclosure level on stock liquidity found in previous studies may be explained by mandatory elements contained in their indexes. Therefore, the voluntary disclosed information is not too useful and valuable for the Tunisian investors to make their decisions. Firms disclose information only for respecting the regulation and still limited.

Indeed, the disclosed information in annual reports is too standard in the sense that all the Tunisian firms provide the same information.

Our results show that there is no valuable information disclosed in web sites. Accordingly, websites' information is not considered a reliable information for Tunisian investors. We conclude that voluntary disclosure does not mitigate information asymmetry. Indeed, in Tunisia as in other Arabic countries (for example Egypt and Jordan), firms do not disclose enough information to investors because of social and cultural characteristics, such as tendency towards secrecy (Hassan et al. 2006; Haddad et al., 2009). For instance, investors do not rely on the firm to obtain useful information they need but prefer paying to collect private information. Consequently, adverse selection risk increases significantly and discourages liquidity traders to negotiate, which decreases stock liquidity. Moreover, we can explain the decisions of Tunisian investors by psychological biases. According to Daniel et al. (1998), investors' overconfidence bias leads to overreaction in the market. Hence, investors' response to public information is limited. The adjustment of investors' decision is too little even if public information contradicts their private information.

Indeed, the survey of Zaiane and Abaoub (2010) confirms that Tunisian investors are overconfident. They find that 66,4% of respondents have confidence in their intuitions while 32,4% of respondents hold their stocks less than three months. Consequently, overconfident investors trade aggressively. Their results show that 55,2% of respondents use more than one source of information (Internet, newspapers and advice of brokers) because they think that they will never get all the hidden information. Hence, they conclude that Tunisian investors overestimate the quality of information and their ability to interpret it.

## 5.4. Robustness tests

### 5.4.1. Bootstrap approach

In small samples, a bootstrap approach might be preferred. This approach consists to simulate new samples obtained by sampling with replacement from the original sample. Results given by this approach are the same found with OLS regression. Hence, we confirm the robustness of previous results.

**Table 7. Relationship between firms' information environment and stock liquidity using Bootstrap approch**

|  | LIUM |
|---|---|
| **AIMO** | 3.296 |
|  | (7.62)** |
| **BOTS** | -89.922 |
|  | (1.05) |
| **SWEB** | -39.697 |
|  | (1.59) |
| **VLAT** | 1.296 |
|  | (0.19) |
| **SIZE** | -2.430 |
|  | (0.58) |
| **BTMK** | 28.667 |
|  | (1.17) |
| **Constant** | -40.623 |
|  | (0.52) |
| **Observations** | 41 |
| **Adjusted R2** | 0.7548 |
| **Legend** | |
| BOTS= disclosure index on annual reports; SWEB= disclosure index on website; AIMO=absolute value of imbalance order; BTMK=book-to-market ratio; SIZE= market capitalization; VLAT=standard deviation daily returns. *, **: statistically significant for the threshold values of 5%, and 1% respectively. | |

### 5.4.2. Other liquidity proxies

To check for robustness of results, we replace the Liu's measure with other liquidity measures. We test whether the previous results depend on the choice of liquidity measures or not. We have two sets of measures capturing two liquidity dimensions: cost and quantity. For assessing the cost of transaction, we rely on 1) bid ask spread (BASQ) frequently used in prior studies[12] as a measure of immediat cost; 2) Amihud illiquidity ratio (ILIQ) which captures the price impact[13] ; and 3) the proportion of zero returns (PZER) which represents the total cost of transaction[14]. In order to measure the transaction volume, we introduce two measures: 1) turnover ratio (TURN) reflecting trading frequency; and 2) market depth[15] (DEPH) employed as a measure of transaction volume. The following table provides results.

**Table 8. Relationship between firms' information environment and other liquidity proxies**

|  | BASQ | PZER | ILIQ | DEPH | TURN |
|---|---|---|---|---|---|
| **AIMO** | 0.021 | 1.115 | 0.057 | -0.019 | -0.064 |
|  | (2.02) | (8.43)** | (4.17)** | (3.85)** | (3.82)** |
| **BOTS** | -3.441 | -42.930 | 0.666 | 4.271 | 0.291 |
|  | (1.50) | (1.25) | (0.19) | (3.26)** | (0.08) |
| **SWEB** | -0.589 | -11.544 | -0.336 | 0.232 | 0.951 |
|  | (1.50) | (1.19) | (0.52) | (0.63) | (1.18) |
| **VLAT** | -0.032 | -0.371 | 0.289 | 0.208 | -0.129 |
|  | (0.29) | (0.14) | (1.48) | (2.08)* | (0.59) |
| **SIZE** | -0.208 | -0.314 | -0.699 | 0.016 | -0.378 |
|  | (2.41)* | (0.16) | (5.57)** | (0.22) | (2.62)* |
| **BTMK** | 0.403 | 14.070 | 1.034 | 0.897 | -1.507 |
|  | (0.70) | (1.89) | (1.52) | (3.18)** | (1.70) |
| **Constant** | 3.918 | -2.928 | 11.805 | 5.909 | 6.745 |
|  | (1.87) | (0.07) | (4.66)** | (3.95)** | (2.22)* |
| Observations | 41 | 41 | 41 | 41 | 41 |
| Adjusted R2 | 0.52 | 0.75 | 0.71 | 0.60 | 0.47 |

**Legend**
BTMK=book-to-market ratio; SIZE= market capitalization; VLAT=standard deviation daily returns; LIUM= standardised turnover-adjusted number of zero daily trading volume; BOTS= disclosure index on annual reports; SWEB= disclosure index on website; AIMO=absolute value of imbalance order.
*, **: statistically significant for the threshold values of 5%, and 1% respectively.

---

[12] Mattoussi et al. (2004); Gana and Chemli (2008) and Haddad et al. (2009).
[13] Espinosa et al. (2008).
[14] Lesmond et al. (1999).
[15] Mattoussi et al. (2004).

Results show that voluntary information disclosure has no effect on transaction cost. Hence, corporate disclosure does not mitigate asymmetric information problems and is not enough to improve stock liquidity. In contrast, we detect positive and significant effect of private information on the measures of transaction costs. One explanation is that when there is an arrival of large number of informed investors into the market, information asymmetry is more severe, consequently transaction costs increase. These results show the robustness of those found using Liu's measure. Moreover, private information reduces the frequency of activity and market depth. In addition, we report that public information published in annual reports improves the market depth. This indicates that information voluntarily disclosed (BOTS) improves the absorption of shares without affecting both frequency and cost of transaction. These results confirm also, that the Tunisian investors do not rely on public information to measure transaction costs.

## 6. Conclusion

In this study, we raised the question of the effect of informational environment of the firm on stock liquidity in Tunisian market, which contains public and private information. Our results show that there is no relationship between private and public information. We find also that only private information influences stock liquidity, and that Tunisian investors do not rely only on information disclosed in annual reports and firms' websites. Contrary to previous empirical findings in emerging market (Mattoussi et al. 2004; Haddad et al., 2009), our study does not support the signaling theory predictions but confirms the predictions of behavioral finance theory. These results may help also to understand the informational environment of Tunisian listed firms. Despite the Tunisian regulation's efforts made to improve the firms' transparency, this is not enough to constrain Tunisian firms to disclose more useful information and to discourage private information collection.

In fact, Tunisian regulators need to incite Tunisian listed companies to disclose more information through fiscal advantages and subventions. In addition, CMF should control the information disseminated and impose penalties when firms did not disclose mandatory information (other than financial statement) in annual report.

Our study presents some limitations. First, we consider a static approach since we consider only firms listed in 2007. Second, we have neglected other sources of information, such as meetings with financial analysts and media representatives. Indeed, this practice has been increasingly adopted by the Tunisian firms as a mean of voluntary disclosure, particularly following the outbreak of the global financial crisis. In 2008, 20 listed companies held 31 meetings with analysts, 17 of which were held during the market downturn because of the financial crisis[16]. Some companies have had more than one meeting in 2008, for example**,** Alkimia has organized 4 meetings. Thus, it would be interesting to see the effect of such new ways of communication on the behavior of Tunisian investors.

---

[16] Tunis Stock Exchange annual report (2008)